\RequirePackage[T1]{fontenc}
\documentclass[conference]{IEEEtran}
\IEEEoverridecommandlockouts
\usepackage{cite}
\usepackage{amsmath,amssymb,amsfonts}
\usepackage{booktabs}
\usepackage{graphicx}
\usepackage{textcomp}
\usepackage[table]{xcolor}
\usepackage{array}
\usepackage{float}
\usepackage{tabularx}
\usepackage{xurl}
\usepackage{tikz}
\definecolor{citationnavy}{RGB}{0,20,115}
\usepackage[colorlinks=true,citecolor=citationnavy,linkcolor=citationnavy,urlcolor=black]{hyperref}
\usetikzlibrary{arrows.meta,calc,positioning}
\definecolor{enactInk}{HTML}{192D42}
\definecolor{enactMuted}{HTML}{526173}
\definecolor{enactBlue}{RGB}{30,69,148}
\definecolor{enactPink}{RGB}{251,185,197}
\definecolor{enactYellow}{RGB}{248,224,88}
\definecolor{enactRed}{RGB}{231,93,64}
\colorlet{enactBlueTint}{enactBlue!7!white}
\colorlet{enactPinkTint}{enactPink!22!white}
\colorlet{enactTeal}{enactPink}
\colorlet{enactTealTint}{enactPinkTint}
\definecolor{enactGray}{HTML}{F3F5F7}
\colorlet{enactOrange}{enactRed}
\colorlet{enactPurple}{enactYellow}
\colorlet{enactPurpleTint}{enactYellow!22!white}
\tikzset{
  enact box/.style={draw=enactInk!55,line width=.65pt,rounded corners=2pt,
    fill=white,align=center,inner sep=5pt,font=\sffamily\fontsize{8.5}{10}\selectfont,text=enactInk},
  enact arrow/.style={-{Latex[length=2mm,width=1.4mm]},draw=enactInk!80,line width=.8pt},
  enact stage/.style={font=\sffamily\bfseries\scriptsize,text=enactMuted,anchor=west},
  enact note/.style={font=\sffamily\scriptsize,text=enactMuted,align=center}
}

\makeatletter
\renewcommand{\subsubsection}{\@startsection{subsubsection}{3}{\parindent}%
  {0ex plus 0.1ex minus 0.1ex}{0ex}{\normalfont\normalsize\bfseries}}
\makeatother

\def\BibTeX{{\rm B\kern-.05em{\sc i\kern-.025em b}\kern-.08em
    T\kern-.1667em\lower.7ex\hbox{E}\kern-.125emX}}

\newcommand{\ENACT}{\textit{ENACT-Heart}~}

\begin{document}

\title{ENACT-Heart -- ENsemble-based Assessment Using CNN and Transformer on Heart Sounds
}

\author{\IEEEauthorblockN{Jiho Han\textsuperscript{1}\thanks{\textsuperscript{1}This work was conducted while Jiho Han was affiliated with the Department of Computer Science and Engineering, The University of Michigan -- Dearborn, Dearborn, MI, USA. He is currently an independent researcher.}}
\IEEEauthorblockA{\textit{Independent Researcher}\\
Seoul, Republic of Korea\\
\href{https://orcid.org/0009-0003-9329-5628}{ORCID: 0009-0003-9329-5628}\\
\href{mailto:jihohan@umich.edu}{jihohan@umich.edu}}
\and
\IEEEauthorblockN{Adnan Shaout}
\IEEEauthorblockA{\textit{Department of Electrical and Computer Engineering} \\
\textit{The University of Michigan -- Dearborn}\\
Dearborn, MI, USA\\
shaout@umich.edu}}

\IEEEaftertitletext{
\begin{minipage}{\textwidth}
  \normalfont\normalsize
  \setlength{\parskip}{0pt}
  \setlength{\abovecaptionskip}{3pt}
  \centering
  \includegraphics[width=\linewidth,trim=0bp 3bp 0bp 0bp,clip]{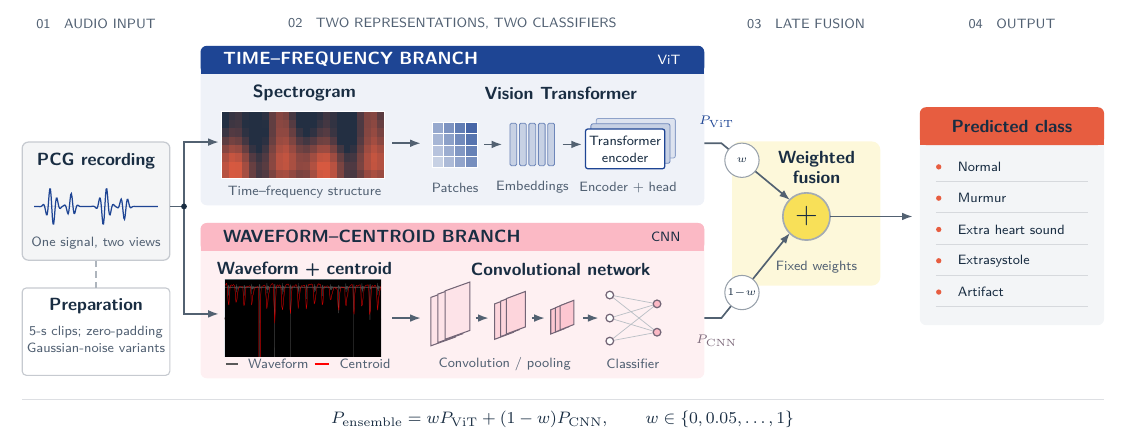}\par
  \nointerlineskip
  \makeatletter
  \def\@captype{figure}
  \makeatother
  \caption{ENACT-Heart combines a spectrogram-based Vision Transformer with a waveform--spectral-centroid CNN through fixed-weight prediction fusion. The waveform--centroid inset shows the gray waveform and red centroid trace. Other signal thumbnails and model glyphs are schematic.}
  \label{fig:banner}
\end{minipage}
\par\vspace{.8\baselineskip}
}
\maketitle

\begin{abstract} \label{abstract}
\textit{ENACT-Heart} is a two-branch ensemble for heart-sound classification. A Vision Transformer (ViT) processes spectrogram images, while a convolutional neural network (CNN) processes waveform--spectral-centroid visualizations derived from the same recordings. The branch predictions are combined by a weighted average with fixed, input-independent weights. On the PASCAL heart-sound dataset, the ensemble achieved an accuracy of 97.52\%, compared with 93.88\% for the ViT and 95.45\% for the CNN. The ensemble achieved a macro-F1 of 0.94 and recall of 0.74 for extra heart sounds. On this dataset, late fusion improved accuracy and macro-F1 relative to both individual branches.
\end{abstract}


\section{Introduction} \label{introduction}
Heart-sound analysis is a non-invasive component of cardiovascular assessment. These sounds, primarily associated with the closure of heart valves, provide information about cardiac function. The first heart sound (S1), produced by the closure of the atrioventricular valves, and the second heart sound (S2), associated with the closure of the semilunar valves, are commonly recognized as the characteristic 'lub' and 'dub.' Analysis of their timing and acoustic characteristics can support the assessment of heart-sound abnormalities.

\begin{table*}[t]
  \centering
  \caption{Related methods and evaluation scope.}
  \label{tab:sota_comp}
  \footnotesize
  \setlength{\tabcolsep}{4pt}
  \renewcommand{\arraystretch}{1.08}
  \begin{tabularx}{\textwidth}{@{}>{\raggedright\arraybackslash}p{2.5cm} >{\raggedright\arraybackslash}p{6.1cm} >{\raggedright\arraybackslash}X@{}}
    \toprule
    Study & Model and fusion strategy & Data and task \\
    \midrule
    Malik et al.~\cite{malik_multi-classification_2022}
      & MFCC features; LSTM classifier
      & PASCAL; PhysioNet 2017. \\
    Yang et al.~\cite{yang_assisting_2023}
      & Transformer classifier
      & Five categories: normal and four valve diseases. \\
    Wang et al.~\cite{wang_pctmf-net_2023}
      & Parallel CNN--Transformer; feature-level fusion
      & Yaseen; PhysioNet 2016, evaluated separately. \\
    Cheng and Sun~\cite{cheng_heart_2023}
      & 1D convolution followed by a Transformer
      & Three datasets; binary and multiclass tasks. \\
    Liu et al.~\cite{liu_heart_2023}
      & Bispectrum features; Vision Transformer
      & PhysioNet 2022; normal/abnormal classification. \\
    Jumphoo et al.~\cite{jumphoo_exploiting_2024}
      & Convolution--DeiT with transfer learning
      & Heart Sound Murmur database; valve-disease classification. \\
    \midrule
    \textbf{ENACT-Heart}
      & Spectrogram ViT + waveform--centroid CNN; fixed-weight prediction-level fusion
      & PASCAL A + B; five labels. \\
    \bottomrule
  \end{tabularx}
  \par\vspace{3pt}
  \begin{minipage}{\textwidth}\footnotesize
    Different datasets, labels and protocols preclude a shared performance ranking. ENACT-Heart's metrics appear in Tables~\ref{tab:perf_summary} and~\ref{tab:perf_comp}.
  \end{minipage}
\end{table*}

\subsection{Anomalies in the Heart Sounds}
Heart murmurs, extra heart sounds, and extrasystoles are common anomalies detected during cardiac auscultation. Each type of anomaly provides valuable information about potential underlying cardiac conditions.

\textbf{Heart murmurs} are among the most common anomalies detected through auscultation. Produced by turbulent blood flow strong enough to generate audible noise, these "whooshing" sounds can be heard in various scenarios, including some healthy individuals. While innocent murmurs, also known as functional or benign murmurs, are typically harmless and not associated with structural heart abnormalities, pathologic murmurs indicate underlying conditions such as valve defects, congenital heart defects, or abnormal blood flow patterns. 

\textbf{Extra Heart Sounds} refer to additional heart sounds beyond the normal "lub-dub" pattern. These may manifest as "lub-lub dub" or "lub dub-dub" sequences. Extra heart sounds can sometimes indicate underlying conditions, although they may also occur in healthy individuals.

\textbf{Extrasystole} involves irregular heart rhythms, typically presenting as extra or skipped heartbeats. These can be heard as sequences such as "lub-lub dub" or "lub dub-dub." While extrasystoles may be benign, they can also signify underlying heart disease, making early detection important for effective treatment.

These distinctions provide context for the five-class task evaluated in this study. The PASCAL dataset \cite{bentley_pascal_nodate} includes murmur, extra heart sound, and extrasystole labels alongside normal and artifact labels. These are recording categories rather than disease diagnoses; their definitions and availability across source subsets are summarized in Table~\ref{tab:heart_anomalies} in Appendix~\ref{app:dataset}.

\subsection{Organization of the Paper}
This paper is organized as follows: Section \ref{background} covers the background of the study, providing the necessary theoretical foundation. \ref{related_works} provides an in-depth review of related work, discussing the advancements and methodologies in cardiac sound analysis and the integration of machine learning models in cardiovascular diagnostics. Section \ref{proposed approach} details the proposed approach, including the data preprocessing techniques, the generation of audiovisual data, and the model architectures. Section \ref{experiments} describes the experimental setup, including the dataset used, training procedures, and evaluation metrics. The results of the experiments are presented in \ref{results}, highlighting the performance of individual models and the ensemble method. Finally, Section \ref{conclusion} concludes the paper with a summary of findings, potential implications for clinical practice, and directions for future research.

\subsection{Contributions}
The contributions of this study are as follows:
\begin{itemize}
    \item We present \ENACT (Fig.~\ref{fig:banner}), a fixed-weight late-fusion ensemble combining a spectrogram-based ViT and a waveform--spectral-centroid CNN for heart-sound classification.
    \item We compare the ensemble with its individual branches on the PASCAL dataset, examining the effect of combining predictions from two representations of the same recording.
    \item We report overall and class-level performance to characterize both the gains from fusion and the variation in classification performance across heart-sound categories.

\end{itemize}

\section{Historical Background} \label{background}
The diagnosis of heart conditions dates back to the early days of medicine, where physicians relied on palpation and pulse assessment to detect abnormalities. A significant breakthrough occurred in the 1700s when Jean Baptiste de Senac, physician to King Louis XV of France, established the connection between atrial fibrillation and mitral valve disease. Senac's work laid the foundation for cardiology as a distinct field of study \cite{mcmichael_history_1982}.

The invention of the stethoscope by René Laennec in 1816 marked a pivotal moment in cardiac diagnostics. Laennec introduced the technique of "mediate auscultation" using his newly created paper acoustic device, allowing for more accurate detection of heart sounds and abnormalities \cite{laennec_treatise_nodate}. This innovation remains a cornerstone in the history of cardiology.

These early diagnostic methods, based on manual interpretation of heart sounds, evolved significantly over the centuries. With technological advancements, traditional auscultation has been augmented by electrocardiograms (ECGs) and other imaging modalities. In recent years, the integration of artificial intelligence (AI) and machine learning has opened new avenues for the analysis of heart sounds, leading to more precise and efficient diagnostic tools \cite{chen_deep_2021, huang_fusion_2020}.

\section{Related Works} \label{related_works}
Table~\ref{tab:sota_comp} summarizes related heart-sound classification methods and their evaluation settings.

\subsection{Thematically Related Works}
Thematically, the proposed method lies in the field of computer-assisted diagnosis (CAD) systems for heart disease. CAD systems for heart diseases utilize computation techniques such as machine learning, pattern recognition, and AI to analyze cardiac data and provide decision-support tools for healthcare providers.

There have been numerous attempts to apply CAD systems for heart diseases in diverse modalities, including but not limited to ECG, cardiac CT/MRI, etc. These systems analyze cardiac data to identify abnormalities and patterns - especially indicators of certain heart diseases. However, the direction of the majority of these researches are pointed mainly toward computer vision over audio AI, mainly due to the advanced deep learning models available.

For some researches that emphasized sound classification, its methodologies have varied slightly from the approach proposed in this study. For instance, Jumphoo et al. utilized a CNN for feature extraction and Data-efficient Image Transformer (DieT), a variant of the ViT model, for classification tasks through stacking \cite{jumphoo_exploiting_2024}. Another heart sound classification model, proposed by Liu et al., also uses ViT for classification but employs a different image modality called bispectral patterns and relies solely on ViT without integrating other models \cite{liu_heart_2023}. While these studies highlight the effectiveness of ViT individually, they do not explore the potential benefits of using an ensemble approach.

\ENACT uses prediction-level fusion between a ViT and a CNN operating on two representations of the same sound input. This describes the design of the present ensemble; the studies reviewed here do not establish that this combination is unprecedented or superior under a shared evaluation protocol.

\subsection{Methodologically Related Works}
The use of computer vision as a tool for machine hearing is an emerging approach. There have been attempts to use computer vision techniques as a method of machine hearing - analyzing audio signals by treating them as visual data. 

Hsu \textit{et al.} \cite{hsu_deep_2021} presented a deep learning-based music classification through mel-spectrogram and Fourier tempogram features. Although the concept of using multiple different audiovisual modalities and models from singular sound data is there, the paper employed the short-chunk CNN + ResNet as the backbone architecture of their models.

\section{Proposed Approach} \label{proposed approach}
This study focuses on classifying heart-sound recordings using two visual representations of the same audio signal. The aim is to evaluate whether combining the predictions of a spectrogram-based ViT and a waveform--spectral-centroid CNN improves classification relative to either branch alone.

Researches has demonstrated that it is
possible to process spectrograms from audio data as images
and apply computer vision algorithms such as CNN \cite{verma_neural_2018, cabrera-ponce_detection_2020, hyder_acoustic_2017}. The core problem of the current approaches in using regular CNN-based computer vision methods on audio spectrogram representation lies in the distinctiveness of the spectrogram in comparison to other image data.

Vision Transformers use self-attention to model relationships between image patches, while CNNs use convolutional filters to extract features. These different processing mechanisms motivate evaluating their predictions together; the comparison in this study does not establish that either architecture is generally superior.

\subsection{Spectral Data Visualization \& Analysis}
In spectral visualization and analysis, researchers employ various techniques to gain insights into the frequency content of signals. These methodologies enable the examination of how frequencies evolve over time, providing valuable information for tasks such as audio processing, speech recognition, and biomedical signal analysis.

\subsubsection{Spectrogram}
Spectrograms stand as one of the primary tools in spectral visualization. They offer detailed representations of frequency spectra over time, revealing how the frequency composition of a signal changes temporally. By plotting frequency on the vertical axis, time on the horizontal axis, and intensity or magnitude using color or brightness, spectrograms provide a comprehensive view of signal dynamics. This detailed visualization allows analysts to identify specific features, patterns, and transient events within the signal, making spectrograms invaluable for tasks requiring fine-grained temporal frequency analysis.

\subsubsection{Spectral Centroid}
In contrast to the detailed time--frequency mapping provided by spectrograms, spectral centroids offer a simplified summary of a signal's frequency content. The spectral centroid indicates the "center of mass" or average frequency of a signal within each time frame. This single-value representation provides a concise summary but does not preserve the full spectral detail available in a spectrogram.

The two representations emphasize different aspects of the recording: the spectrogram depicts its time--frequency structure, while the waveform--centroid visualization combines time-domain amplitude with a frame-level spectral summary. This motivates their use as potentially complementary inputs to the two classifiers.

The proposed ensemble combines predictions from the two branches rather than their intermediate features. The branches differ in both architecture and input representation; the available comparison does not isolate the contribution of either choice.

\subsection{Fixed-Weight Prediction Fusion}
Let $P_{\mathrm{ViT}}$ and $P_{\mathrm{CNN}}$ denote the branch prediction vectors in the same class order. The ensemble forms the weighted average
\begin{equation}
\label{eq:late_fusion}
P_{\mathrm{ensemble}} = wP_{\mathrm{ViT}} + (1-w)P_{\mathrm{CNN}},
\end{equation}
where a single scalar weight $w$ is shared across inputs. This is prediction-level late fusion with fixed weights rather than input-dependent routing.

Weight combinations were searched from 0 to 1 in increments of 0.05, as detailed in Section~\ref{experiments}.

Figure~\ref{fig:banner} shows the two branches and their fixed-weight fusion. Figure~\ref{fig:flowchart} summarizes the data-preparation operations.

\section{Experiments} \label{experiments}
In this section, we describe the experimental setup. The data-preparation operations are summarized in Fig.~\ref{fig:flowchart}; the prediction pipeline is shown separately in Fig.~\ref{fig:banner}.

\subsection{Dataset Used}
The heart sound dataset provided in the PASCAL Classifying Heart Sounds Challenge was used for training and testing the models \cite{bentley_pascal_nodate}. The audio files in the dataset vary in length, ranging from 1 second to 30 seconds. The five classification labels are normal, murmur, extra heart sound, extrasystole, and artifact; their definitions and source subsets are given in Table~\ref{tab:heart_anomalies} in Appendix~\ref{app:dataset}.

\begin{figure}[t]
\centering
\resizebox{\columnwidth}{!}{\begin{tikzpicture}[x=1cm,y=1cm]
  \path[use as bounding box] (-.1,-.9) rectangle (8.6,6.1);
  \node[enact stage] at (0,5.85) {DATA-PREPARATION OPERATIONS};
  \node[enact box,text width=7.75cm,minimum height=.85cm] (raw) at (4.25,5.05)
    {\textbf{PASCAL heart-sound recordings}\\[2pt]\scriptsize Subsets A and B pooled; five labels};
  \node[enact box,text width=7.75cm,minimum height=.85cm,below=4.5mm of raw] (clips)
    {\textbf{Clip preparation}\\[2pt]\scriptsize Segment into 5-s clips; zero-pad shorter clips};
  \node[enact box,text width=7.75cm,minimum height=1.0cm,below=4.5mm of clips] (aug)
    {\textbf{Gaussian-noise augmentation}\\[2pt]\scriptsize Nine noisy versions per original; $\mu=0$, $\sigma=0.1$};
  \node[enact box,draw=enactBlue,fill=enactBlueTint,text width=3.65cm,minimum height=1.4cm,anchor=north] (spec) at ($(aug.south)+(-2.125,-.9)$)
    {\textbf{Spectrogram image}\\[2pt]\scriptsize Low-pass cutoff: 195 Hz};
  \node[enact box,draw=enactTeal,fill=enactTealTint,text width=3.65cm,minimum height=1.4cm,anchor=north] (wave) at ($(aug.south)+(2.125,-.9)$)
    {\textbf{Waveform + centroid}\\[2pt]\scriptsize Normalized waveform\\with spectral centroid};
  \draw[enact arrow,shorten <=1pt,shorten >=1pt] (raw.south) -- (clips.north);
  \draw[enact arrow,shorten <=1pt,shorten >=1pt] (clips.south) -- (aug.north);
  \coordinate (branch-junction) at ($(aug.south)+(0,-.45)$);
  \draw[draw=enactInk!80,line width=.8pt,shorten <=1pt] (aug.south) -- (branch-junction);
  \draw[enact arrow,shorten >=1pt] (branch-junction) -| (spec.north);
  \draw[enact arrow,shorten >=1pt] (branch-junction) -| (wave.north);
\end{tikzpicture}}
\caption{Data preparation for the two model branches: clip-length standardization, Gaussian-noise augmentation, and generation of spectrogram and waveform--spectral-centroid images.}
\label{fig:flowchart}
\end{figure}

\subsection{Data Preprocessing}
\subsubsection{Creating the Dataframe}
The PASCAL dataset consists of two main folders, each containing labeled audio files. These folders were combined into a single dataframe, which includes the file paths and corresponding labels. An exploratory data analysis (EDA) was performed to understand the distribution of classes and identify any potential issues such as missing or mislabeled data.

\subsubsection{Clip-Length Standardization}
To address the inconsistent length of audio recordings, a preprocessing step was implemented. Each audio file was segmented into 5-second clips. If an audio file was shorter than 5 seconds, it was zero-padded to meet the required input length.

\subsubsection{Data Augmentation}
To increase the dataset size and introduce variability, Gaussian noise was added to the audio files, generating 9 additional noisy versions for each original file. Gaussian noise with a mean of 0 and a standard deviation of 0.1 was added to the audio files to generate augmented data. Each audio file was segmented into 5-second clips to standardize the input length for model training. This resulted in a total of 10 versions per audio file (1 original + 9 augmented). The augmented audio data was included in the dataframe, and corresponding labels were updated to reflect the augmentation process.

\subsection{Audiovisual Image Data Generation}
Two types of visual representations were generated for each audio file: spectrograms and centroid graphs.

\subsubsection{Spectrogram}
Spectrograms were generated to visualize the frequency content of the audio files over time. Given the presence of various background noises in real-world conditions, a low-pass filter set at 195 Hz was applied. This filter helps to emphasize the cardiac sounds, which predominantly occur in the lower frequency range, while reducing noise from higher frequencies. The spectrogram images were saved with dimensions that matched the input requirements of the models.

\subsubsection{Waveform--Spectral-Centroid Visualization}
The spectral centroid, representing the "center of mass" of the spectrum, was calculated for each audio file. This metric summarizes the distribution of spectral energy within each frame. A centroid graph was generated, overlaying the normalized waveform and centroid values. This combined visualization served as the input to the CNN model. The rationale for this representation was twofold:
\begin{enumerate}
    \item The waveform depicts the signal's amplitude variation over time.
    \item The spectral centroid adds a frame-level summary of frequency content, providing another view of the recording alongside the waveform.
\end{enumerate}

\begin{table*}[t]
  \centering
  \caption{Overall classification performance on the pooled PASCAL dataset. Boldface denotes the highest value for each metric.}
  \label{tab:perf_summary}
  \small
  \setlength{\tabcolsep}{5pt}
  \renewcommand{\arraystretch}{1.08}
  \begin{tabular*}{\textwidth}{@{\extracolsep{\fill}}l r ccc ccc@{}}
    \toprule
    & & \multicolumn{3}{c}{Macro average} & \multicolumn{3}{c}{Weighted average} \\
    \cmidrule(lr){3-5}\cmidrule(l){6-8}
    Model & Accuracy (\%) & Precision & Recall & F1 & Precision & Recall & F1 \\
    \midrule
    ViT & 93.88 & 0.89 & 0.88 & 0.88 & 0.94 & 0.94 & 0.94 \\
    CNN & 95.45 & 0.94 & 0.89 & 0.91 & 0.95 & 0.95 & 0.95 \\
    ENACT-Heart & \textbf{97.52} & \textbf{0.96} & \textbf{0.92} & \textbf{0.94} & \textbf{0.97} & \textbf{0.98} & \textbf{0.97} \\
    \bottomrule
  \end{tabular*}
\end{table*}

\begin{table*}[t]
  \centering
  \caption{Class-level classification performance on the pooled PASCAL dataset. Boldface denotes the highest value across models for each class and metric.}
  \label{tab:perf_comp}
  \small
  \setlength{\tabcolsep}{5pt}
  \renewcommand{\arraystretch}{1.08}
  \begin{tabular*}{\textwidth}{@{\extracolsep{\fill}}l ccc ccc ccc@{}}
    \toprule
    & \multicolumn{3}{c}{ViT} & \multicolumn{3}{c}{CNN} & \multicolumn{3}{c}{ENACT-Heart} \\
    \cmidrule(lr){2-4}\cmidrule(lr){5-7}\cmidrule(l){8-10}
    Class & Precision & Recall & F1 & Precision & Recall & F1 & Precision & Recall & F1 \\
    \midrule
    Artifact & 0.80 & 0.96 & 0.87 & 0.92 & 0.93 & 0.92 & \textbf{0.93} & \textbf{0.97} & \textbf{0.95} \\
    Extra heart sound & 0.73 & 0.65 & 0.69 & 0.86 & 0.67 & 0.76 & \textbf{0.89} & \textbf{0.74} & \textbf{0.81} \\
    Extrasystole & 0.98 & 0.84 & 0.91 & 0.98 & 0.87 & 0.92 & \textbf{1.00} & \textbf{0.91} & \textbf{0.96} \\
    Murmur & 0.94 & 0.97 & 0.96 & 0.95 & 0.97 & 0.96 & \textbf{0.98} & \textbf{0.99} & \textbf{0.99} \\
    Normal & \textbf{0.99} & 0.95 & 0.97 & 0.97 & 0.98 & 0.98 & 0.98 & \textbf{0.99} & \textbf{0.99} \\
    \bottomrule
  \end{tabular*}
\end{table*}

\subsection{Model Training}
Both branches were trained without pretrained weights.

\subsubsection{ViT}
The ViT model was trained using the generated spectrogram images. The training process involved splitting the data into training and validation sets, followed by model training with appropriate hyperparameters such as batch size of 32 and 50 epochs, using an Adam optimizer with a learning rate of 0.001. Data augmentation techniques, such as random noise addition, were applied to improve model robustness.

\subsubsection{CNN}
The CNN model was trained using the centroid graphs. The CNN model architecture included three convolutional layers followed by max-pooling and dropout layers. The data was split into training and validation sets, and the model was trained with optimized hyperparameters, including a batch size of 32 and 50 epochs, using an Adam optimizer with a learning rate of 0.001. The CNN model also used the augmented data generated during preprocessing.

\subsection{Ensemble Method}
The ensemble combined the ViT and CNN predictions according to Eq.~\eqref{eq:late_fusion}. Weight combinations were systematically tested on the grid
\begin{equation}
\label{eq:weight_grid}
\begin{aligned}
w_{\mathrm{ViT}} &= w = 0.05k, \\
w_{\mathrm{CNN}} &= 1-w, \\
k &\in \{0,1,\ldots,20\}.
\end{aligned}
\end{equation}
Thus, the ViT weight ranged from 0 to 1 in increments of 0.05, with the CNN receiving the complementary weight. These weights were global scalars, not outputs of a sample-dependent gating network.

\section{Results} \label{results}

\subsection{Individual Model Performance}

The models classified recordings into five dataset categories: artifact, extra heart sound, extrasystole, murmur, and normal. Table~\ref{tab:perf_summary} separates overall accuracy from macro-averaged and weighted metrics; Table~\ref{tab:perf_comp} reports the class-level results.

\subsubsection{ViT} The spectrogram-based ViT achieved 93.88\% accuracy and a macro-F1 of 0.88. Its lowest recall was 0.65 for extra heart sounds.

\subsubsection{CNN} The waveform--centroid CNN achieved 95.45\% accuracy and a macro-F1 of 0.91. Both the architecture and the input representation differ between the branches, so these results alone do not isolate the contribution of either factor.

\subsection{Ensemble Model Performance}

The ensemble combines the branch predictions using fixed weights, as shown in Fig.~\ref{fig:banner}. Weight combinations were evaluated using the grid described in Section~\ref{experiments}.

The ensemble achieved an accuracy of 97.52\%, a difference of 2.07 percentage points relative to the CNN. Macro-F1 was 0.94 and macro-recall was 0.92. Extra-heart-sound recall was 0.74, illustrating why accuracy alone does not describe performance across classes. Weighted precision, recall and F1 were 0.97, 0.98 and 0.97, respectively. The confusion-matrix panels are provided in Fig.~\ref{fig:conf_mat} in Appendix~\ref{app:confusion}.

\subsection{Context in Prior Work}
Table~\ref{tab:sota_comp} summarizes the approaches and evaluation settings of selected related studies. These studies use different datasets, label definitions and evaluation protocols; their published accuracy values therefore do not constitute a shared benchmark against which \ENACT can be ranked.

The table distinguishes prediction-level fusion in \ENACT from the parallel feature fusion of PCTMF-Net \cite{wang_pctmf-net_2023} and the convolution--Transformer architectures of other studies. It provides methodological context rather than evidence of state-of-the-art performance.

\section{Conclusion and Discussion} \label{conclusion}
\ENACT combines a spectrogram-based ViT and a waveform--spectral-centroid CNN through fixed-weight late fusion. On the PASCAL dataset, the ensemble achieved 97.52\% accuracy, compared with 93.88\% for the ViT and 95.45\% for the CNN, and a macro-F1 of 0.94. In this evaluation, combining the two branches improved accuracy and macro-F1 relative to either branch alone.

The gains vary across classes and metrics. In particular, recall for extra heart sounds was 0.74, indicating scope for improving recognition of this category. Because both the architecture and the input representation differ between the branches, further ablation studies would help distinguish their individual contributions. Similarly, evaluating augmentation separately would help establish its effect on robustness rather than attributing the ensemble's gains to augmentation alone.

Portable or wearable heart-sound analysis is a possible application for future investigation. Before such use, the approach would require evaluation with independent recordings, different acquisition devices and realistic background noise, as well as assessment of computational cost and latency. Its effects on diagnostic decisions, healthcare costs and patient outcomes would require separate clinical and implementation studies.

A broader design implication is the flexibility to pair different preprocessing methods with different classifiers. Because late fusion operates on predictions for a shared set of classes, the branches need not use matching internal feature representations. Expressing the audio as images also provides a route to explore compatible pretrained vision models in future work. Neither branch used pretrained weights in these experiments, so the benefit of transferring previously learned features remains to be evaluated. The two-branch results motivate further investigation of this modular approach, including alternative preprocessing methods and model backbones.

{
    \footnotesize
    \IEEEtriggeratref{7}
    \bibliographystyle{IEEEtran}
    \bibliography{refs}
}

\onecolumn
\appendices
\section{Dataset Labels and Source Subsets}
\label{app:dataset}

\setcounter{table}{0}
\setcounter{figure}{0}
\renewcommand{\thetable}{A.\Roman{table}}
\renewcommand{\thefigure}{B.\arabic{figure}}
\renewcommand{\theHtable}{appendix.\arabic{table}}
\renewcommand{\theHfigure}{appendix.\arabic{figure}}

Table~\ref{tab:heart_anomalies} lists the dataset labels and their source subsets.
\begin{table}[H]
  \centering
  \begin{minipage}{.88\linewidth}
  \centering
  \caption{Label definitions and source subsets in PASCAL CHSC2011 \cite{bentley_pascal_nodate}.}
  \label{tab:heart_anomalies}
  \small
  \setlength{\tabcolsep}{4pt}
  \renewcommand{\arraystretch}{1.05}
  \begin{tabularx}{\linewidth}{@{}>{\raggedright\arraybackslash}p{2.7cm} c >{\raggedright\arraybackslash}X@{}}
    \toprule
    Label & Subset & Dataset description \\
    \midrule
    Normal & A, B & Recognizable S1--S2 pattern. \\
    Murmur & A, B & Murmur-like sound between principal sounds. \\
    Extra heart sound & A & Additional sound beyond S1--S2. \\
    Extrasystole & B & Rhythm disturbance; extra or skipped beat. \\
    Artifact & A & Non-cardiac sound or interference. \\
    \bottomrule
  \end{tabularx}
  \par\vspace{3pt}
  \begin{minipage}{\columnwidth}\footnotesize
    A: iStethoscope Pro; B: DigiScope. These are dataset labels, not disease diagnoses. Label availability differs by source.
  \end{minipage}
  \end{minipage}
\end{table}

\section{Confusion Matrices}
\label{app:confusion}
Figure~\ref{fig:conf_mat} presents the confusion-matrix panels for normal,
murmur, extrasystole, and extra heart sound classifications.
\begin{figure}[H]
    \centering
    \begin{minipage}{\linewidth}
    \input{figures/confusion_matrix_layout}
    \end{minipage}
    \par\vspace{4pt}
    \begin{minipage}{.88\linewidth}
    \caption{Confusion-matrix panels for four heart-sound categories. Each panel uses its own color scale. The figure does not include an artifact panel, numerical cell annotations or prediction-axis labels.}
    \label{fig:conf_mat}
    \end{minipage}
\end{figure}

\section{Change Log}
\label{app:changelog}

\noindent\textbf{Version 1 (2025-02-24):} Initial arXiv version.

\par\smallskip
\noindent\textbf{Version 2 (2026-11-04):} Revised terminology, equations, figures, tables, and discussion; corrected a reference, reorganized appendices, and updated author information. No new experiments.

\end{document}